\documentclass[11pt]{article}
\usepackage{graphicx}
\usepackage{placeins}
\usepackage{float}
\usepackage{subfigure}
\usepackage[toc,page]{appendix}

\usepackage{amssymb}
\usepackage{epsfig}
\begin{document}
\title{{\bf{\Large Near horizon hidden symmetry and entropy of Sultana-Dyer black hole: A time dependent case}}}
\author{ 
 {\bf {\normalsize Bibhas Ranjan Majhi}$
$\thanks{E-mail: bibhas.majhi@iitg.ernet.in}}\\ 
{\normalsize Department of Physics, Indian Institute of Technology Guwahati,}
\\{\normalsize Guwahati 781039, Assam, India}
\\[0.3cm]
}

\maketitle

\begin{abstract}
  The non-extremal stationary black holes do have ``{\it hidden}'' conformal symmetry in which case the generators form $SL(2,\mathbb{R})$ group. Here, I explicitly show that the Sultana-Dyer black hole solution also possesses the similar near horizon conformal symmetry. This is the first example of time dependent case which also exhibits such feature. Moreover, using the corresponding generators I find the expression of the horizon entropy in the context of Virasoro algebra and Cardy formula. The result matches with the earlier findings. The present analysis is important since in reality the metric is not stationary and hence one must look into these realistic situations to obtain more information about our universe. I expect that the analysis will illuminate certain features of the time dependent situations.    
\end{abstract}

\section{Introduction}
   The geometry of an extremal Kerr black hole, in the near horizon limit, has $SL(2,\mathbb{R})\times U(1)$ group symmetry \cite{Bardeen:1999px}. Following the observation that the certain non-trivial diffeomorphisms of the near horizon extremal Kerr (NHEK) geometry lead to two copies of Virasoro algebra whose central charge gives the entropy, it has been conjectured in \cite{Guica:2008mu} that extremal Kerr black hole is dual to two dimensional conformal field theory (CFT). Unfortunately, all attempts to extend the analysis for non-extremal cases has been failed. Later on, a different approach, based on the wave equation for a massless scalar field in the non-extremal Kerr background, has been proposed \cite{Castro:2010fd}. It was found that the radial part of the scalar equation in the near horizon and low frequency limits exhibits $SL(2,\mathbb{R})_L\times SL(2,\mathbb{R})_R$ symmetry. This has been extended in various cases \cite{Matsuo:2010in,Bertini:2011ga}. 

     Till now, so far I am aware of, none of the existing discussions encounter the time dependent situation. The main obstacle is that the scalar field can not be decomposed in the mode functions in which the time part is of the form $\sim e^{-i\omega t}$ because the metric coefficients do not have the time translational symmetry. Due to this the rest of the programme can not be simply followed. Now the question is why do we need to worry about such situations? The answer is as follows. In our practical life the black holes are a part of the cosmological model and they are surrounded by a local mass distribution. Therefore, the metrics are not asymptotically flat and moreover the spacetimes evolve with time. So the stationary results are no longer blindly acceptable without a prior verification. Therefore, in order to know the real universe one needs to study the time dependent black hole solutions. The aim of the paper is two fold. First we shall study whether a time dependent solution possesses any near horizon $SL(2,\mathbb{R})$ symmetry like the stationary ones. Finally, it will be investigated whether the generators of the symmetry can illuminate the horizon entropy. 

  It must be pointed out that the straightforward application of the existing method \cite{Castro:2010fd} is not feasible. This is because the analysis is mainly based on that fact that the stationary metric is time independent. Therefore there exists a timelike Killing vector, corresponding to which the energy of any particle on this spacetime is conserved. But in the evolving cases we do not have such vector. Hence the mode decomposition, as stated in the above paragraph, is not possible. Therefore, in this paper I shall adopt a different approach based on the solutions of the conformal Killing equation (CKE) near the horizon introduced earlier in \cite{Franzin:2011wi} for a static spherically symmetric metric. Since no introduction of the auxiliary scalar field is needed here, one can, in principle, use this method for any spacetime metric. The steps are as follows. Solution of the CKE for a specific submanifold of spacetime, particularly the ($t-r$) sector, near the horizon leads to a set of diffeomorphism vectors. Then it can be shown that the generators corresponding to them form $SL(2,\mathbb{R})$ algebra.

   In this paper, I shall consider this alternative approach to uncover the hidden symmetries of the Sultana-Dyer (SD) black hole \cite{Sultana:2005tp}. The metric is related to the Schwarzschild metric by a time dependent conformal factor. Also it a solution of general relativity with two types of fluid as a source. I will show that the solution of CKE leads to generators which a identical to those of the Schwarzschild metric. Therefore, following the arguments made in \cite{Franzin:2011wi}, we can immediately conclude that the SD has conformal symmetry near the horizon which follows the $SL(2,\mathbb{R})$ group. So far I know, this will be the first example for the evolving case which exhibits such phenomenon. 
Next issue I shall address in this paper is the role played by these generators in finding the entropy of the horizon. There are some attempts for the stationary case \cite{Matsuo:2010in} but they are not complete and free of ambiguity. Here I shall first find the near horizon, regular form of the generators and the calculate a bracket among the Noether charges, obtained earlier in one of my papers \cite{Majhi:2011ws} (See also \cite{Majhi:2012tf,Majhi:2013lba,Majhi:2012nq,Majhi:2014lka}){\footnote{The original approach was done in \cite{Brown:1986nw} and the complete literature for further development can be followed from \cite{Majhi:2012st}.}}. The central charge corresponding to the eigenmodes $+1$ and $-1$ will lead to the correct value of the entropy upon using in the Cardy formula \cite{Cardy:1986ie,Bloete:1986qm,Carlip:1998qw} which matches with the earlier findings \cite{Majhi:2014lka,Majhi:2014hpa}.    
     
  I shall organise the paper as follows. In the next section, a brief introduction on SD metric will be presented. In section \ref{generators}, the near horizon conformal generators for the submanifold will be found out. Also it will be shown that they satisfy the $SL(2,\mathbb{R})$ algebra. Then using them the entropy of the horizon will be calculated in section \ref{Cardy}. Finally, I shall conclude. For clarity and completeness, three appendices, containing some details of the analysis, will be given at the end. 
Before going into the next discussion let me now introduce the notations which I shall adopt in this paper. The unbar quantities correspond to the seed metric while the bar ones are for conformal metric. The small Latin indices $a,b,c$, etc. represent all the spacetime indices and the large ones, like $A,B,C,$ etc. denote the angular (or transverse) coordinates.    

\section{\label{SD}SD metric: a brief review}
   In this section, the Sultana-Dyer (SD) spacetime will be introduced with some salient features.
   The SD metric is a cosmological black hole solution of GR with two noninteracting perfect fluids: one is timelike and the other one is null-like. It is an expanding black hole in the asymptotic background of the Einstein-de Sitter universe. The spacetime is asymptotically FLRW. One can obtain this by just giving a time dependent conformal transformation of the Schwarzschild black hole metric (Details can be seen from the original paper of Sultana and Dyer \cite{Sultana:2005tp}). The metric is given by \cite{Sultana:2005tp}:
\begin{equation}
\bar{ds}^2 = a^2(\eta)\Big[-d\eta^2+dr^2+r^2(d\theta^2+\sin^2\theta d\phi^2)+\frac{2M}{r}(d\eta+dr)^2\Big]~.
\label{SD1} 
\end{equation}
The constant $M$, taken to be positive, is identified as the mass of the Schwarzschild black hole and $a^2(\eta)$ is the conformal factor whose explicit form is $a(\eta) = \eta^2$. Here $\eta$, $r$ are the time and the radial coordinates, respectively while $\theta$ and $\phi$ are angular coordinates. Later on I shall, in general, denote these transverse (or angular) coordinates as $x_{\perp}$. By imposing the coordinate transformation $\eta = t + 2M \ln(r/2M - 1)$ in (\ref{SD1}), we can express the SD metric in Schwarzschild like coordinates:
\begin{equation}
\bar{ds}^2 =  a^2(t,r)\Big[-F(r)dt^2 + \frac{dr^2}{F(r)}+ r^2(d\theta^2
+\sin^2\theta d\phi^2)\Big]~,
\label{SDSC}
\end{equation}
where $F(r) = 1-2M/r$.
The conformal factor in this case turns out to be \cite{Faraoni:2013aba}
\begin{equation}
a(t,r) = \Big(t+2M\ln\Big|\frac{r}{2M}-1\Big|\Big)^2~.
\label{a}
\end{equation}
Note that the above diverges near the event horizon $r=2M$ of the Schwarzschild spacetime.

  Let us now find the location of the horizon. To obtain this we shall present below a general analysis following \cite{Jacobson:1993pf}. First denote the stationary seed metric as $g_{ab}$. Remember that for the present case this is actually the Schwarzschild spacetime. Since the seed metric is time independent, there must exists a timelike Killing vector $\chi^a$ such that $\pounds_{\chi}g_{ab}=0$ and the Killing horizon is determined by $g_{ab}\chi^a\chi^b\equiv\chi^2=0$. 
Next consider a metric $\bar{g}_{ab}$ such that it is related to the earlier one by a spacetime dependent conformal factor; i.e. $\bar{g}_{ab}=\Omega^2g_{ab}$. In this case one can show that \cite{Jacobson:1993pf}
\begin{equation}
\pounds_{\bar{\chi}}\bar{g}_{ab} = (\pounds_{\bar{\chi}}\Omega^2) g_{ab} = (\pounds_{\bar{\chi}}\ln\Omega^2)
\bar{g}_{ab}~,
\label{2.01} 
\end{equation}
where $\pounds_{\bar{\chi}}$ is the Lie derivative along $\bar{\chi}^a$; i.e. $\bar{\chi}^a=\chi^a$ is the conformal Killing vector of the metric $\bar{g}_{ab}$. Although the contravariant components are identical to that of the seed metric, the covariant components are related to that of seed by the conformal factor: $\bar{\chi}_a = \bar{g}_{ab}\bar{\chi}^b=\Omega^2g_{ab}\chi^b=\Omega^2\chi_a$. Now as $\chi^2=0$ at the Killing horizon, we must have $\bar{g}_{ab}\bar{\chi}^a\bar{\chi}^b=\Omega^2\chi^2 =0$, provided $\Omega^2\neq\mathcal{O}(\chi^{-2})$. This implies that the same remains as the horizon for the conformal metric. We shall call the later one as the {\it conformal Killing horizon}.
Therefore, $r=2M$ is the conformal horizon of SD metric. In the present paper, we shall define all the quantities on this horizon.

     Before going into the main discussion, I shall conclude this section by pointing out a issue of the SD solution. 
It is known that the SD metric is a solution in GR gravity sourced by two types of matter distribution: one is timelike dust and other is null fluid. The energy-momentum tensor of this theory is $T^{ab} = \mu u^au^b + \tau k^ak^b$, where the first term is for timelike dust with energy density $\mu$ and zero pressure while last term is for null source. Here, the four velocity, $u^a$ is timelike and $k^a$ is the null vector. The non-trivial components of $u^a$ and $k^a$ are the time and radial components. The explicit expressions of them are given by \cite{Sultana:2005tp},
\begin{eqnarray}
&&u^0 = \frac{r^2+M(2r-\eta)}{r\eta^2\sqrt{r^2+2M(r-\eta)}}~; \,\,\,\ u^1 = \frac{M(\eta-2r)}{r\eta^2\sqrt{r^2+2M(r-\eta)}}~;
\nonumber
\\
&& k^0 = \frac{\sqrt{2M(r-\eta)+r^2}}{r\eta^2}~; \,\,\,\ k^1 = - \frac{\sqrt{2M(r-\eta)+r^2}}{r\eta^2}~.
\label{uk}
\end{eqnarray}
It is observed in \cite{Sultana:2005tp} that the energy density of the dust $\mu = \frac{12(r^2+2M(r-\eta))}{r^2\eta^6}$ is positive in the region
\begin{equation}
\eta<\frac{r(r+2M)}{2M}~.
\label{eta}
\end{equation}
Note that in this region of $\eta$ both $u^0$ and $k^0$ are positive while both $u^1$ and $k^1$ are negative when we are on the horizon $r=2M$. So the energy conditions are satisfied and it implies that an observer who is outside the horizon, will see that the dust and the null fluid flow radially into the black hole. On the other hand, for late times; i.e. for $\eta>\frac{r(r+2M)}{2M}$ where all the energy conditions are broken, the sources become unphysical and the dust becomes superluminal. For $r=2M$, the physically acceptable region of time coordinate is $\eta<4M$. Although there exists such a unphysical feature in the SD solution, it is still interesting to study the different aspects as the global structure is similar to that of a cosmological black hole and therefore represents a more realistic situation. Moreover, the metric is simple enough (as it is conformally related to the simplest solution, Schwarzschild spacetime, of GR) to handle and so as a starting example we can study it to explore more information about our universe (The calculation of different thermodynamical quantities for SD metric has been done in \cite{Faraoni:2007gq,Saida:2007ru,Faraoni:2014lsa,Majhi:2014lka,Majhi:2014hpa}.). Of course, we feel that this unphysical superluminal feature is due to the fact that it is far from the realistic one. However, it is expected that if one can find an exact solution of this theory, that will be free of this problem. In absence of such solutions, here I shall adopt the SD metric as a model, representing our ``real world'', with the expectation that it mimics the realistic situation.

\section{\label{generators}Near horizon symmetry and the generators}
     Here, by solving the conformal Killing equation under the SD background, I shall show that the generators corresponding to conformal Killing vectors, in the near horizon limit, obey the standard $SL(2,\mathbb{R})$ algebra. The analysis will be similar to that of \cite{Franzin:2011wi}. Throughout this paper, I define the horizon by the location of the radial coordinate where the norm of the conformal Killing vector vanishes. As discussed in the earlier section (see below Eq. (\ref{2.01})), the horizon, in this case, is given by $r=2M$. The term ``near horizon limit'' here will be used to imply that at the end of the every calculation $r\rightarrow 2M$ limit will be taken.   

   To start with, consider $\xi^a$ as the conformal Killing vector of the seed metric $g_{ab}$; i.e. $\pounds_{\xi}g_{ab} = f(x)g_{ab}$ where $f(x)$ is a function of space-time coordinates. Therefore for any diffeomorphism $x^a\rightarrow x^a + \bar{\xi^a} = x^a + \xi^a$ of the conformal metric $\bar{g}_{ab}=\Omega^2g_{ab}$, we obtain
\begin{equation}
\pounds_{\bar{\xi}}\bar{g}_{ab} = \pounds_{\xi}(\Omega^2 g_{ab}) = (\pounds_{\bar{\xi}}\ln \Omega^2 + f)\bar{g}_{ab}~,
\label{3.01}
\end{equation}
i.e. $\bar{\xi}^a = \xi^a$ is also a conformal Killing vector for the metric $\bar{g}_{ab}$, but in this case the proportionality factor on the right hand side of the equation is shifted by $\pounds_{\bar{\xi}}\ln \Omega^2$. This is the consequence of the earlier discussion around Eq. (\ref{2.01}) with the difference that in the later case the diffeomorphism is the conformal Killing vector of the seed metric, instead of the Killing vector.

    Next we want to solve (\ref{3.01}) to find the vectors for the SD metric (\ref{SDSC}), like as has been done in \cite{Franzin:2011wi}, under the assumptions $\bar{\xi}^t\equiv\bar{\xi}^t(t,r)$, $\bar{\xi}^r\equiv\bar{\xi}^r(t,r)$ and $\bar{\xi}^A = 0$ where $A$ corresponds to the transverse indices. This choice is motivated by the fact that the near horizon theory is a two dimensional conformal theory and the metric of a black hole is effectively given by a two dimensional form in which only the ($t-r$) sector is important \cite{Majhi:2011yi}. That means the mass and interaction terms of the action of the external field do not contribute and the theory becomes a two dimensional free one. The SD spacetime is also behaves in a similar manner. An explicit analysis has already been done in Appendix of \cite{Majhi:2014hpa}. Keeping this in mind and since our whole analysis is near to the horizon one can make such a restricted diffeomorphism which only deforms the ($t-r$) sector. 
Also remember that here $\Omega=a(t,r)$, which is given by (\ref{a}).  
With these assumptions let us first consider the angular part of the equation (\ref{3.01}); i.e. $a=A$ and $b=B$. Then the choice $A=\theta=B$ leads to
\begin{equation}
f(x) = \frac{2F}{a^2r}\bar{\xi}_r~.
\label{sigma}
\end{equation}
An identical result can also be obtained for $A=\phi=B$. This helps us to find the factor $f$ in terms of the metric coefficients and the conformal Killing vector. We shall use it in the following calculation to find the non-trivial expressions for the components of $\bar{\xi}^a$.
Expansion of (\ref{3.01}) under the background (\ref{SDSC}) and use of (\ref{sigma}) lead to the following non-trivial equations:
\begin{eqnarray}
&&\frac{\partial}{\partial t}\Big(\frac{\bar{\xi}_t}{a^2}\Big) - \frac{FF'}{2a^2}\bar{\xi}_r  = -\frac{F^2}{a^2r}\bar{\xi}_r~;
\label{3.02}
\\
&&\frac{\partial}{\partial r}\Big(\frac{\bar{\xi}_r}{a^2}\Big)+\frac{F'}{2a^2F}\bar{\xi}_r = \frac{1}{a^2r}\bar{\xi}_r~;
\label{3.03}
\\
&&\frac{\partial}{\partial t}\Big(\frac{\bar{\xi}_r}{a^2}\Big) + \frac{\partial}{\partial r}\Big(\frac{\bar{\xi}_t}{a^2}\Big) - \frac{F'}{a^2F}\bar{\xi}_t = 0~.
\label{3.04}
\end{eqnarray}
The prime denotes the derivative with respect to the radial coordinate ``$r$''. In the above, the first equation is for the choice $a=t=b$ while the second one is for $a=r=b$ and the last one comes by taking $a=t, b=r$. Others are trivially satisfied.

  In principle the solutions of the above equations give us the required components. Here I shall solve them in the near horizon limit; i.e. $r\rightarrow 2M$. In this limit, remember that $F$ vanishes and $a$ diverges while $a^2F\rightarrow 0$. Therefore, the terms on the right hand sides of equations (\ref{3.02}) and (\ref{3.03}) can be neglected as they are decreasing fast compared to the other terms. Hence the above equations simplify to the following forms in the near horizon region:
\begin{eqnarray}
&&\frac{\partial}{\partial t}\Big(\frac{\bar{\xi}_t}{a^2}\Big) - \frac{FF'}{2a^2}\bar{\xi}_r  = 0~;
\label{3.05}
\\
&&\frac{\partial}{\partial r}\Big(\frac{\bar{\xi}_r}{a^2}\Big)+\frac{F'}{2a^2F}\bar{\xi}_r = 0~;
\label{3.06}
\\
&&\frac{\partial}{\partial t}\Big(\frac{\bar{\xi}_r}{a^2}\Big) + \frac{\partial}{\partial r}\Big(\frac{\bar{\xi}_t}{a^2}\Big) - \frac{F'}{a^2F}\bar{\xi}_t = 0~.
\label{3.07}
\end{eqnarray}
The above can be solved by using the identical steps employed in \cite{Franzin:2011wi}. The solutions can be directly written by borrowing the general logic presented below Eq. (\ref{2.01}). Since, the terms on the right hand side of Eqs. (\ref{3.02}) and (\ref{3.03}) do not contribute in the near horizon limit, the equations (\ref{3.05}) -- (\ref{3.07}) represents the conformal Killing equations which is the modification of the Killing equations of Schwarzschild metric by a conformal factor. Therefore, as stated above, the contravariant components for SD metric is identical to those of Schwarzschild spacetime while the covariant components will be modified by a conformal factor which, in the present case, is given by $a^2$. Therefore, as the components for the Schwarzschild metric have already been obtained in \cite{Franzin:2011wi}, we can write those for SD metric directly and hence the generators. But for the completeness and as the paper should be self-contained so that a new reader can understand, I present below some details for obtaining the components by solving Eqs. (\ref{3.05}) -- (\ref{3.07}). 
The solutions are
\begin{eqnarray}
&&\bar{\xi}_t = a^2 \Big[K F + \frac{F'\sqrt{F}}{2\sqrt{\lambda}}\Big(\alpha e^{\sqrt{\lambda}~t} - \beta e^{-\sqrt{\lambda}~t}\Big)\Big]~;
\label{3.08}
\\
&&\bar{\xi}_r = \frac{a^2}{\sqrt{F}}\Big(\alpha e^{\sqrt{\lambda}~t} + \beta e^{-\sqrt{\lambda}~t}\Big)~.
\label{3.09}
\end{eqnarray}
Here $K$, $\alpha$, $\beta$ and $\lambda$ are integration constants. For completeness, I present the steps to obtain these in the Appendix \ref{AppA}.
It must be noted that for the present case the covariant components of the vector are modified by the conformal factor $a^2$ compared to the static case which was obtained earlier in \cite{Franzin:2011wi}. Also since these are calculated near the horizon, where the $f(x)$ factor has been neglected (see Eq. (\ref{sigma})), the static case vector reduces to the Killing vector while that for the SD metric remains conformal Killing vector. This is happening as the term $\pounds_{\bar{\xi}}\ln\Omega^2$ in (\ref{3.01}) contributes to the near horizon equations (\ref{3.05})--(\ref{3.07}). This fact is the consequence of the general argument given around Eq. (\ref{2.01}).

   Although the covariant components get modified by the conformal factor, the contravariant components remain identical to the static metric. This is because to find them one has to raise the index by the inverse conformal metric $\bar{g}^{ab} = \Omega^{-2}g^{ab} = a^{-2}g^{ab}$. Below I give the expressions for the contravariant components: 
\begin{eqnarray}
&&\bar{\xi}^t = -K-\frac{F'}{2\sqrt{\lambda F}}\Big(\alpha e^{\sqrt{\lambda}~t} - \beta e^{-\sqrt{\lambda}~t}\Big)
\nonumber
\\
&&\bar{\xi}^r = \sqrt{F}\Big(\alpha e^{\sqrt{\lambda}~t} + \beta e^{-\sqrt{\lambda}~t}\Big)~.
\label{3.10}
\end{eqnarray} 
The generators corresponding these can be found by the different choice the constants $K,\alpha$ and $\beta$.  
Note that the components, given in (\ref{3.10}),  are identical to what were obtained in \cite{Franzin:2011wi} for the static case. Therefore, the generators will also be same for the SD metric. Therefore here I do not give the steps to achieve the final forms. These can be followed from \cite{Franzin:2011wi}. The near horizon generators are given by 
\begin{eqnarray}
&&{\bar{H}}_{+1} = {H}_{+1}=i\gamma e^{\kappa t}\Big(\sqrt{F}\partial_r - \frac{F'}{2\kappa\sqrt{F}}\partial_t\Big)~;
\nonumber
\\
&&\bar{H}_0 = H_0 = -\frac{i}{\kappa}\partial_t~;
\nonumber
\\
&&\bar{H}_{-1} = H_{-1} = - i\gamma e^{-\kappa t}\Big(\sqrt{F}\partial_r + \frac{F'}{2\kappa\sqrt{F}}\partial_t\Big)~,
\label{3.11}
\end{eqnarray}
where $\kappa = F'(2M)/2$ is the surface gravity of the seed metric and $\gamma$ is a constant.
One can notice that these satisfy the $SL(2,\mathbb{R})$ algebra
\begin{equation}
[\bar{H}_0,\bar{H}_{\pm 1}]= \mp i \bar{H}_{\pm 1}; \,\,\,\,\ [\bar{H}_{+1},\bar{H}_{-1}] = 2i\bar{H}_0~,
\label{SL2R}
\end{equation}
with the choice $\gamma^2 F''(r) =2$. A brief details to obtain the above form of the generators is also given in Appendix \ref{AppB}.
 This tells that the above is valid for the value of $F(r)$ upto order $(r-2M)^2$ (For a detailed analysis, see \cite{Franzin:2011wi}).

   So what I found so far is that the generators corresponding to the  near horizon conformal Killing vectors, obtained by solving conformal Killing equation under certain conditions, obey $SL(2,\mathbb{R})$ algebra where the metric coefficient $F(r)$ has to be upto $\mathcal{O}(r^2)$. This implies that the near horizon SD metric has a conformal symmetry like the static Schwarzschild \cite{Bertini:2011ga,Franzin:2011wi} or stationary Kerr \cite{Castro:2010fd} solutions. So far I am aware of, this is the first example of time dependent case which exhibits the ``hidden'' conformal symmetry.

\section{\label{Cardy}Virasoro generators and entropy}
   After finding the ``hidden'' symmetries, I am interested to examine if they have any role in the thermodynamics of horizon. This issue will be addressed in the present section in the context of Virasoro algebra and Cardy formula. First I shall find the convenient Virasoro generators and then calculate the charge defined in my earlier papers \cite{Majhi:2011ws,Majhi:2014lka}. Finally this will be compared with the standard algebra to identify the correct zero mode eigenvalue and central charge to obtain the entropy of the horizon.

  To obtain the Virasoro generators, let us first rewrite (\ref{3.11}) in the following form. In the near horizon we substitute $F'(r\rightarrow 2M)=2\kappa$. This will lead to
\begin{eqnarray}
&&{\bar{H}}_{+1} = i\gamma e^{\kappa t}\frac{1}{\sqrt{F}}\Big(F\partial_r - \partial_t\Big)~;
\nonumber
\\
&&{\bar{H}}_0 = -\frac{i}{\kappa}\partial_t~;
\nonumber
\\
&&\bar{H}_{-1} = - i\gamma e^{-\kappa t}\frac{1}{\sqrt{F}}\Big(F\partial_r + \partial_t\Big)~,
\label{4.01}
\end{eqnarray}
Next express the above in the null coordinates, given by
\begin{equation}
u=t-r^*; \,\,\,\,\ v=t+r^*~,
\label{4.02}
\end{equation}
where $r^*$ is the tortoise coordinate, defined through the differential equation $dr^* = dr/F$. In this coordinates (\ref{4.01}) reduces to the following form:
\begin{eqnarray}
&&{\bar{H}}_{+1} = -\frac{2i}{\kappa\sqrt{F}} e^{\frac{\kappa}{2}(u+v)}\partial_u~;
\nonumber
\\
&&\bar{H}_0  = -\frac{i}{\kappa}(\partial_u+\partial_v)~;
\nonumber
\\
&&\bar{H}_{-1} = -\frac{2i}{\kappa\sqrt{F}} e^{-\frac{\kappa}{2}(u+v)}\partial_v~,
\label{4.03}
\end{eqnarray}
with the choice $\gamma=1/\kappa$.
Notice that $\bar{H}_{+1}$ and $\bar{H}_{-1}$ are not regular at the horizon. Here I shall choose the Virasoro generators such that the components must be regular near the horizon so that all the derived results are finite in this limit. For that consider the following redefined general form of the generators:
\begin{equation}
\bar{H}_n=-\frac{i}{\kappa}e^{\frac{n\kappa}{2}(u+v)}\Big[(1+n)\partial_u + (1-n)\partial_v\Big]~,
\label{4.04}
\end{equation}
with $n=+1,0,-1$. In the above $\bar{H}_{+1}$ and $\bar{H}_{-1}$ have been scaled by the factor $\sqrt{F}$. So the components of the parameter, corresponding to the generator (\ref{4.04}) can be read off as
\begin{equation}
\bar{\zeta}_n^u = -\frac{i(1+n)}{\kappa}e^{\frac{n\kappa}{2}(u+v)}; \,\,\,\ \bar{\zeta}_n^v = -\frac{i(1-n)}{\kappa}e^{\frac{n\kappa}{2}(u+v)}~.
\label{4.05}
\end{equation}
Next impose the following properties on the Virasoro generators: (i) the zero mode value of the Noether charge corresponding this vector, calculated at the horizon, is real, (ii) the Lie bracket among them obeys one sub-algebra isomorphic to Diff $S^1$; i.e. $i\{\xi_m,\xi_n\}^a = (m-n)\xi_{m+n}^a$ and (iii) they are periodic function of time coordinate $t=(u+v)/2$ as Euclidean time has a periodicity $2\pi/\kappa$. These can be achieved by diving by $i$ and then replacing $n$ by $-in$ in the right hand side of the above.
So we choose the components of Virasoro generators as
\begin{equation}
\bar{\xi}_n^u = -\frac{(1-in)}{\kappa}e^{\frac{-in\kappa}{2}(u+v)}; \,\,\,\ \bar{\xi}_n^v = -\frac{(1+in)}{\kappa}e^{\frac{-in\kappa}{2}(u+v)}~.
\label{4.06}
\end{equation}
Remember that contravariant components are same both for the seed metric and the conformal metric while covariant components among these spacetimes are differ by the conformal factor. So the above are also the Virasoro generators for the Schwarzschild metric; i.e. $\bar{\xi}^a_n = \xi^a_n$.

   Now I shall calculate the Noether charge and the bracket among the charges corresponding to diffeomorphism near the horizon. Here (\ref{4.06}) will be chosen as the diffeomorphism vector. The charge and the bracket for the conformal metric are given by \cite{Majhi:2011ws}
\begin{equation}
\bar{Q}_m = \frac{1}{2}\int_{\mathcal{H}}\sqrt{\bar{\sigma}} d\bar{\Sigma}_{ab}\bar{J}^{ab}[\bar{\xi}_m]~,
\label{Qm}
\end{equation} 
and 
\begin{equation}
[\bar{Q}_m,\bar{Q}_n]=\int_{\mathcal{H}} \sqrt{\bar{\sigma}} d\bar{\Sigma}_{ab}\Big[\bar{\xi}^a_n\bar{J}^b_m - (m\leftrightarrow n)\Big]~,
\label{QmQn}
\end{equation}
respectively.  Here $\bar{\sigma}$ is the determinant of the reduced metric on the horizon and the potential, which is an anti-symmetric tensor, is in the following form:
\begin{equation}
\bar{J}^{ab}[\bar{\xi}_m] = \frac{1}{16\pi G}\Big(\nabla^a\bar{\xi}^b_m - \nabla^b\bar{\xi}^a_m\Big)~,
\label{4.09}
\end{equation}
where $\bar{J}^a_m \equiv \bar{J}^a[\bar{\xi}_m]$ and so on. The Noether current is defined as $\bar{J}^a_m = \bar{\nabla}_b\bar{J}^{ab}_m$.
The surface element on the horizon is $d\bar{\Sigma}_{ab} = -d^2x_{\perp}(\bar{N}_a\bar{M}_b - \bar{N}_b\bar{M}_a)$. $\bar{N}^a$ and $\bar{M}^a$ are the spacelike and timelike unit normals, respectively which satisfy $\bar{g}_{ab}\bar{N}^a\bar{N}^b = +1$ and $\bar{g}_{ab}\bar{M}^a\bar{M}^b=-1$. For clarity, a brief discussion has been given in Appendix \ref{AppC}.
For simplicity of the calculation, I shall express (\ref{Qm}) and (\ref{QmQn}) in terms of the quantities defined for the seed metric. This has already been done by me in one of my earlier papers \cite{Majhi:2014lka}. The expressions are as follows:
\begin{equation}
\bar{Q}_m = \frac{1}{2}\int_{\mathcal{H}}\sqrt{\sigma}~\Omega^2 d\Sigma_{ab}\Big[J^{ab}[\xi_m]+\frac{2}{16\pi G}\xi^b_m\nabla^a(\ln \Omega^2)\Big]~,
\label{4.07}
\end{equation} 
and 
\begin{equation}
[\bar{Q}_m,\bar{Q}_n]=\int_{\mathcal{H}} \sqrt{\sigma}~\Omega^2 d\Sigma_{ab}\Big[(\xi^a_nJ^b_m+\xi^a_n K^b_m)- (m\leftrightarrow n)\Big]~.
\label{4.08}
\end{equation}
Here $K^b_m$ is given by
\begin{eqnarray}
K^b_m = \frac{2}{\Omega}J^{bc}[\xi_m]\nabla_c\Omega - \frac{1}{\Omega^2}\nabla_c K^{bc}_m~,
\label{4.10}
\end{eqnarray}
where $K^{ab}_m=1/16\pi G(\xi^a_m\nabla_b\Omega^2 - \xi^b_m\nabla_a\Omega^2)$. The above expressions are valid for any diffiomorphism vector for which the relation between the conformal and seed vectors is $\bar{\xi}^a=\xi^a$, $\bar{\xi}_a=\Omega^2\xi_a$. Since the similar happens for (\ref{4.06}), the above charge and bracket expressions can also be used here. In the next these will be calculated explicitly by using (\ref{4.06}) as they are also applicable to seed metric.

  For the Schwarzschild metric, in $(t,r,x_{\perp})$ coordinate system, the unit normals are given by $N^a=(0,\sqrt{F},{\bf 0})$ and $M^a=(-1/\sqrt{F},0,{\bf 0})$. Transforming them in light cone coordinates (\ref{4.02}) we find 
\begin{equation} 
N^a= (-\frac{1}{\sqrt{F}},\frac{1}{\sqrt{F}},{\bf{0}}); \,\,\,\ M^{a} =  (-\frac{1}{\sqrt{F}},-\frac{1}{\sqrt{F}},{\bf{0}})~.
\label{4.11}
\end{equation}
The above are written in the order $(u,v,x_\perp)$. So to calculate (\ref{4.07}) and (\ref{4.08}) explicitly, we need only the $uv$ component of the surface element. This turns out to be $d\Sigma^{uv} = (-2/F)d^2x_{\perp}$. Now using this and substituting the expressions for Virasoro generators (\ref{4.06}) in the first and the last terms of (\ref{4.07}), we find 
\begin{eqnarray}
d\Sigma^{ab}J_{ab}[\xi_m] = 2d\Sigma^{uv}J_{uv} = d^2x_{\perp}e^{-\frac{im\kappa}{2}(u+v)}\Big(\frac{2F'}{\kappa}+2m^2\Big)~,
\label{4.12}
\end{eqnarray}
and 
\begin{eqnarray}
d\Sigma^{ab}\xi_{mb}\nabla_a(\ln\Omega^2) &=& d\Sigma^{uv}\Big[\xi_{mu}\nabla_v(\ln a^2)- \xi_{mv}\nabla_u(\ln a^2)\Big]
\nonumber
\\
&=& -d^2x_{\perp} \frac{2}{\kappa}e^{-\frac{im\kappa}{2}(u+v)}\Big(\frac{Fa'}{a}+\frac{im\dot{a}}{a}\Big)~,
\label{4.13}
\end{eqnarray}
respectively. Since $a$ is given by (\ref{a}), it is easy to see $a'=(4Ma^{1/2})/(rF)$ while $\dot{a} = 2a^{1/2}$. Therefore the term within the bracket in (\ref{4.13}) is of the order $a^{-1/2}$ which vanishes near the horizon as $a$ diverges in this limit. Hence the last term in (\ref{4.07}) can be neglected compared to the first term. Then substituting (\ref{4.12}) in (\ref{4.07}) and integrating near the horizon we obtain
\begin{equation}
\bar{Q}_m = \frac{\bar{A}}{8\pi G}(1+m^2/2)e^{-\frac{im\kappa}{2}(u+v)}~, 
\label{4.14}
\end{equation}
where $\bar{A}=16\pi M^2a^2$ is the horizon area of the SD metric. Similarly, one can show that the term $d\Sigma_{ab}\xi^a_nK^b_m$ in the bracket (\ref{4.08}) will not contribute near the horizon. So in an identical way one finds
\begin{eqnarray}
[\bar{Q}_m,\bar{Q}_n] &=&-i(m-n)\Big[\bar{Q}_{m+n} - \frac{\bar{A}}{16\pi G}(m+n)^2e^{-\frac{i(m+n)\kappa}{2}(u+v)}\Big] 
\nonumber
\\
&-& i\frac{\bar{A}}{16\pi G}(m^3-n^3)e^{-\frac{i(m+n)\kappa}{2}(u+v)}~.
\label{4.15}
\end{eqnarray}
Therefore the central term comes out to be
\begin{equation}
\bar{K}[\bar{\xi}_m,\bar{\xi}_n] = [\bar{Q}_m,\bar{Q}_n] + i(m-n)\bar{Q}_{m+n}= -i\frac{\bar{A}}{16\pi G}e^{-\frac{i(m+n)\kappa}{2}(u+v)}(mn^2-m^2n)~.
\label{4.16}
\end{equation}  
Now in the usual two dimensional conformal field theory, the central term is given by $-im^3(\bar{C}/12)\delta_{m+n,0}$, where $\bar{C}$ is the central charge \cite{DiFrancesco:1997nk}. Therefore from the above we find
\begin{equation}
\bar{K}[\bar{\xi}_{+1},\bar{\xi}_{-1}] = -i\frac{\bar{A}}{8\pi G}\equiv -i\frac{\bar{C}}{12}~.
\label{4.17}
\end{equation}
Similarly, taking $m=0$ in (\ref{4.14}) one obtains the zero mode eigenvalue of the charge. Hence our values of $\bar{Q}_0$ and $\bar{C}$ are 
\begin{equation}
\bar{Q}_0 = \frac{\bar{A}}{8\pi G}; \,\,\,\ \frac{\bar{C}}{12} = \frac{\bar{A}}{8\pi G}~.
\label{QC}
\end{equation}
Finally, the Cardy formula \cite{Cardy:1986ie}--\cite{Carlip:1998qw} yields the entropy as
\begin{equation}
\bar{S} = 2\pi\sqrt{\frac{\bar{C}(\bar{Q}_0-\bar{C}/24)}{6}} = \frac{\bar{A}}{4G}~.
\label{entropy}
\end{equation} 
The identical value was assumed earlier in \cite{Faraoni:2007gq,Majhi:2014hpa} and later derived in \cite{Majhi:2014lka}.

 Before closing the present section, let me make couple of comments. The Virasoro generators (\ref{4.06}) were constructed by imposing certain conditions on the original ones (\ref{4.03}) which led to the correct expression for the entropy. Particularly, when (\ref{4.04}) was constructed, we rescaled the original ones by factor $\sqrt{F}$ to make them regular near the horizon. Hence, it is expected that (\ref{4.04}) are no longer the solutions of the near horizon conformal Killing equations (\ref{3.05}) -- (\ref{3.07}). One can check that these equations are not even satisfied upto $\mathcal{O}(\bar{\chi}^2)$; i.e. the generators are not conformal Killing vectors even at very near to the horizon. Still it is interesting to verify that they represent the $SL(2,\mathbb{R})$ algebra. That means, these regular ones correspond to the states of some conformal field theory (CFT); or in other words, the near horizon quantum states can be identified with those of CFT. 
Now the question is: What type of symmetry is represented by the vectors (\ref{4.06})? For that let us first calculate the quantity $\bar{\xi}^a\bar{\xi}^b\pounds_{\bar{\xi}}\bar{g}_{ab}$. It turns out that it is of the order $\bar{\chi}^2$ which vanishes near the horizon. This implies that now the Virasoro generators satisfy more weaker condition than (\ref{3.01}). Hence the new vectors correspond to asymptotic symmetry of the particular boundary condition $\bar{\xi}^a\bar{\xi}^b\pounds_{\bar{\xi}}\bar{g}_{ab}=\mathcal{O}(\bar{\chi}^2)$.

  Finally, it must be noted that the modes of the charge (\ref{4.14}) and the central term (\ref{4.16}), hence the central charge, are not finite in the near horizon limit as the conformal factor $a$ diverges (see Eq. (\ref{a})). This is due to the bad choice of coordinates. In Schwarzschild like coordinates $a$ diverges while in ($\eta,r,\theta,\phi$) coordinates $a$ is finite as $a=\eta^2$. On the other hand, when we calculated the charges, the Virasoro generators (\ref{4.05}) were obtained by regularising (\ref{4.03}).  The reason is to obtain the finite results. This is necessary as (\ref{Qm}) and (\ref{QmQn}) are invariant in any coordinate systems and so whatever coordinate we choose, the final result will be same. Hence if we do not take the regular one, our final results are always divergent, irrespective of any coordinate system, even for the usual Schwarzschild metric. Therefore, the regularisation in the generators are necessary.

\section{\label{Conclusions}Conclusions}
   It has been generously agreed that the near horizon non-extremal stationary black hole geometry exhibits a ``hidden'' conformal symmetry. This has been explored by studying the massless Klein-Gordon equation under the metric background. The radial equation in the low frequency limit, near the horizon, is generated by the $SL(2,\mathbb{R})$ group generators \cite{Guica:2008mu}. This has been explored in a completely different way in \cite{Franzin:2011wi} by studying the conformal Killing equation for the submanifold, mainly the ($t-r$) sector, of a static spherically symmetric metric. In this paper, I took time dependent black hole solution and examined if it has similar symmetry. The metric represents a solution of GR with two types of fluid as sources, known as SD spacetime \cite{Sultana:2005tp}. Since it does not has time symmetry the original approach is not applicable. Here I followed the second method.

    I found that the generators, coming from the conformal Killing equation near the horizon for the ($t-r$) sector, exhibit $SL(2,\mathbb{R})$ symmetry. So far I know, this is the first instance for evolving case where such issue has been addressed. This is an important attempt as in reality one does not encounter stationary black holes. Next I calculated a bracket among the Noether charges, defined earlier in one of my papers \cite{Majhi:2011ws}, for these generators. It was observed that for the $+1$ and $-1$ modes the bracket gives the correct value of the central charge which leads to entropy of the horizon. This expression matches with the earlier result \cite{Majhi:2014lka,Majhi:2014hpa} obtained for the SD metric. 

  Now let me conclude by mentioning the following weakness of the present paper which need to be further investigated. First of all, the SD solution is not the exact representation of our real universe. It is a prototype example and hence it has some basic issues which I have already mentioned in the second section. I believe the exact solution will be free of these flaws and in lack of this, the SD metric can enlighten several features of time dependent situation. Thats why this study is important. Furthermore, the correct Virasoro generators were obtained by regularizing the original ones which has been done by hand. Interestingly, the new ones also satisfy the $SL(2,\mathbb{R})$ algebra. That means the near horizon of SD metric has the conformal symmetry. It would have been nice if there exists a direct method to obtain the generators and the ``hidden'' symmetry which will lead to all the results in more concrete way. Finally, it is worth to point out that although the Schwarzschild/CFT correspondence is motivated by the Kerr/CFT correspondence, originally introduced in \cite{Castro:2010fd}; but the structure of generators for both the cases does not have any connection. For instance, one can not reach to Schwarzschild/CFT generators by just taking the limit $J\rightarrow 0$ of those for the Kerr/CFT case, where $J$ is the angular momentum of Kerr black hole. It is quite surprising. Further study in this direction will be very interesting by itself.  All these are under investigations.

\vskip 9mm
\section*{Acknowledgments}
The research of the author is supported by a START-UP RESEARCH GRANT (No. SG/PHY/P/BRM/01) from Indian Institute of Technology Guwahati, India.

\section*{Appendices}
\appendix
\section{\label{AppA}Evaluation of Equations (\ref{3.08}) and (\ref{3.09})}
\renewcommand{\theequation}{A.\arabic{equation}}
\setcounter{equation}{0} 
First solve Eq. (\ref{3.06}). The solution is
\begin{equation}
\frac{\bar{\xi}_r}{a^2}=\frac{A(t)}{\sqrt{F}}~,
\label{A1}
\end{equation}
where $A(t)$ is the integration constant. Next substitute this in (\ref{3.05}) and (\ref{3.07}), respectively. That leads to
\begin{eqnarray}
&&\frac{\partial}{\partial t}\Big(\frac{\bar{\xi}_t}{a^2}\Big)-\frac{F'\sqrt{F}}{2}A(t)=0~;
\label{A2}
\\
&&\frac{\dot{A}}{\sqrt{F}}+\frac{\partial}{\partial r}\Big(\frac{\bar{\xi}_t}{a^2}\Big)-\frac{F'}{a^2F}\bar{\xi}_t = 0~.
\label{A3}
\end{eqnarray}
Taking time derivative of (\ref{A3}) and then using (\ref{A2}) we obtain
\begin{equation}
\ddot{A}+\frac{1}{4}\Big(2FF''-F'^2\Big)A=0~.
\label{A4}
\end{equation}
As $A$ is function of time only while $F$ is function of radial coordinate, the above equation can be separated into time dependent part and radial coordinate dependent part. Both of them must be equal to a constant, say $\lambda$. Then the two equations are as follows:
\begin{eqnarray}
&& \ddot{A}-\lambda A=0~;
\label{A5}
\\
&&2FF'' - (F')^2 + 4\lambda = 0~.
\label{A6}
\end{eqnarray}
The solution for $A$ is
\begin{equation}
A(t) = \alpha e^{\sqrt{\lambda}~t}+\beta e^{-\sqrt{\lambda}~t}~,
\label{A7}
\end{equation}
where $\alpha$ and $\beta$ are two integration constants. Substitution of it in (\ref{A1}) leads to (\ref{3.09}). 
Now to find $\bar{\xi}_t$, we integrate (\ref{A2}) first:
\begin{equation}
\frac{\bar{\xi}_t}{a^2} = \frac{F'\sqrt{F}}{2}\int A(t)dt + g(r)~.
\label{A8}
\end{equation}
Here $g(r)$ is the constant of integration. Using (\ref{A5}) in the above and then absorbing the part which depends only on radial coordinate in $g(r)$ we obtain
\begin{equation}
\frac{\bar{\xi}_t}{a^2} = \frac{F'\sqrt{F}}{2\lambda} \dot{A}(t) + g(r)~.
\label{A9}
\end{equation}
Substitution of the value of $\bar{\xi}_t/{a^2}$ from the above in (\ref{A3}) and finally use of (\ref{A6}) lead to
\begin{equation}
\frac{g'(r)}{g(r)}=\frac{F'}{F}~.
\label{A10}
\end{equation} 
Solution of it is
\begin{equation}
g(r)=KF(r)~,
\label{A11}
\end{equation}
where $K$ is the constant of integration. Making use of the value of $A(t)$ from (\ref{A7}) and $g(r)$ from (\ref{A11}) in (\ref{A9}) we find (\ref{3.08}).

\section{\label{AppB}Derivation of generators (\ref{3.11})}
\renewcommand{\theequation}{B.\arabic{equation}}
\setcounter{equation}{0}
  Eq. (\ref{3.10}) has three independent constants. Three choices of the values of them will lead to three generators. The choices are as follows: (i) $\alpha=i$, $\beta=0=K$; (ii) $\alpha=0=\beta$, $K\neq 0$; and (iii) $\beta=-i$, $\alpha=0=K$. Correspondingly, the generators are:
\begin{eqnarray}
&&{\bar{\tilde{H}}}_{+1} =i e^{\sqrt{\lambda}~ t}\Big(\sqrt{F}\partial_r - \frac{F'}{2\sqrt{\lambda F}}\partial_t\Big)~;
\nonumber
\\
&&\bar{\tilde{H}}_0  = -K\partial_t~;
\nonumber
\\
&&\bar{\tilde{H}}_{-1} = - i e^{-\sqrt{\lambda}~ t}\Big(\sqrt{F}\partial_r + \frac{F'}{2\sqrt{\lambda F}}\partial_t\Big)~,
\label{B1}
\end{eqnarray}
Now to achieve the $SL(2,\mathbb{R})$ algebra we redefine the above as $\bar{H}_0 = \bar{\tilde{H}}_0$ and ${\bar{H}}_{\pm 1} = \gamma {\bar{\tilde{H}}}_{\pm 1}$. Then the commutators are given by (\ref{SL2R}) with the choices: $K\sqrt{\lambda}=i$ and $\gamma^2 F'' = 2$. Solution of the second choice leads to
\begin{equation}
F(r) = \frac{r^2}{\gamma^2}+C_0r + C_1~,
\label{B2}
\end{equation}
where $C_0$ and $C_1$ are integration constants. This tells that the results are valid near the horizon upto order $r^2$. The constants can be determined by imposing the conditions $F(r_H)=0$ and $F'(r_H)=2\kappa$ where $r_H$ is the location of the horizon and $\kappa$ is the surface gravity. These lead to $C_0 = 2\kappa-2r_H/\gamma^2$ and $C_1=r_H^2/\gamma^2-2\kappa r_H$. Using all these in (\ref{A6}) we find $\lambda=\kappa^2$ and so $K=i/\sqrt{\lambda}=i/\kappa$. Substituting them in (\ref{B1}) we obtain the final expressions for generators presented in Eq. (\ref{3.11}).

\section{\label{AppC}Noether current, potential and a bracket among the charges}
\renewcommand{\theequation}{C.\arabic{equation}}
\setcounter{equation}{0} 
Here I shall use the Noether prescription to obtain the current and potential for a gravity theory. This will be based on the Lagrangian formalism. 
Consider a covariant Lagrangian $L(\bar{g}_{ab}, \bar{R}_{abcd})$. Under the metric variation, it varies as
\begin{equation}
\delta(L\sqrt{-\bar{g}})= \sqrt{-\bar{g}}\Big[\bar{E}_{ab}\delta\bar{g}^{ab}+\bar{\nabla}_a\delta v^a\Big]~.
\label{new1}
\end{equation}
The first term corresponds to the equation of motion while the last term is the total derivative term and so it is a surface contribution. Its explicit form turns out to be $\delta v^j = 2\bar{P}^{ibjd}(\bar{\nabla}_b \delta\bar{g}_{di})-2\delta\bar{g}_{di}(\bar{\nabla}_c\bar{P}^{ijcd})$ where $\bar{P}^{abcd}=\partial L/\partial \bar{R}_{abcd}$. In this paper we are interested on the variation due to a diffeomorphism $x^a\rightarrow x^a+\bar{\xi}^a$, in which case $\delta \bar{g}_{ab}=\bar{\nabla}_a\bar{\xi}_b+\bar{\nabla}_a\bar{\xi}_b$. In this case the right hand side of the above can be expressed as total derivative term: $\sqrt{-\bar{g}}\Big[\bar{\nabla}_a(-2\bar{E}^a_b\bar{\xi}^b+\delta v^a)\Big]$. On the other hand, as the Lagrangian is a scalar, the Lie variation of it is given by $\delta(L\sqrt{-\bar{g}}) = \sqrt{-\bar{g}}\bar{\nabla}_a(L\bar{\xi}^a)$. Equating these two one finds a conservation relation $\bar{\nabla}_a\bar{J}^a=0$, where $\bar{J}^a$ is the Noether current. Here it is given by
\begin{equation}
\bar{J}^a = L\bar{\xi}^a+2\bar{E}^{ab}\bar{\xi}_b-\delta v^a~.
\label{new2}
\end{equation}
Note that the above is valid for any diffeomorphism vector $\bar{\xi}^a$. Now as $\bar{J}^a$ is conserved, it can be expressed as covariant derivative of an anti-symmetric tensor $\bar{J}^{ab}$; i.e. $\bar{J}^{a}=\bar{\nabla}_b\bar{J}^{ab}$. In literature, $\bar{J}^{ab}$ is called as the Noether potential.
In the case of the covariant Lagrangian, these are coming out to be \cite{Paddy}
\begin{eqnarray}
\bar{J}^{a}&=&2\bar{P}^{abcd}\bar{\nabla} _{b}\bar{\nabla} _{c}\bar{\xi} _{d}-2\bar{\nabla} _{b}
\left(\bar{P}^{adbc}+\bar{P}^{acbd}\right)\bar{\nabla} _{c}\bar{\xi} _{d}-4\bar{\xi} _{d}\bar{\nabla} _{b}\bar{\nabla} _{c}\bar{P}^{abcd}~;
\label{vir08}
\\
\bar{J}^{ab}&=&2\bar{P}^{abcd}\bar{\nabla} _{c}\bar{\xi} _{d}-4\left(\bar{\nabla} _{c}\bar{P}^{abcd} \right)\bar{\xi} _{d}~,
\label{vir09}
\end{eqnarray}
For GR theory, substitution of each term in the above leads to $\bar{J}^a = \bar{\nabla}_b\bar{J}^{ab}$ where $\bar{J}^{ab}=\bar{\nabla}^a\bar{\xi}^b - \bar{\nabla}^b\bar{\xi}^a$. For the details, see the discussion given in page $394$ of the book \cite{Paddy}. Remember that the expressions for current and potential are general and can be used for any diffeomorphism vector. Inserting the proper normalisation factor $1/16\pi G$ and integrating over the horizon we obtain (\ref{Qm}). 

  To define a bracket among the charges, we take another variation of the charge $\bar{Q}[\bar{\xi}_m]$ for diffemorphism $x^a\rightarrow x^a+\bar{\xi}^a_n$. Then an anti-symmetric combination of $m,n$ is taken which is in the following form:
\begin{equation}
[\bar{Q}_m,\bar{Q}_n]\equiv \delta_{\bar{\xi}_m}\bar{Q}_n - \delta_{\bar{\xi}_n}\bar{Q}_m~.
\label{C5}
\end{equation}
This leads to Eq. (\ref{QmQn}) (For details, see \cite{Majhi:2011ws}).
For GR theory, this exactly matches with the result obtained in \cite{Carlip:1999cy} by Hamiltonian formalism.

\end{document}